\documentclass[prd,twocolumn,twoside,preprintnumbers,superscriptaddress,nofootinbib,natbib,bm,amsrefs]{revtex4-2}

\usepackage{graphicx}
\usepackage{url}
\usepackage{xcolor}
\usepackage{amsmath,amstext,amsfonts,amssymb,amsthm}
\usepackage{comment}
\usepackage[utf8]{inputenc}
\usepackage{multirow}

\renewcommand{\arraystretch}{1.3}

\definecolor{BlueViolet}{rgb}{0.2,0.00,0.7}
\definecolor{Blue}{rgb}{0.15,0.00,0.9}
\definecolor{lightblue}{rgb}{0.15,0.35,0.95}
\definecolor{kitgreen}{rgb}{0,0.58823,0.50980}
\usepackage[
 colorlinks=true,
 linkcolor=lightblue,
 citecolor=lightblue,
 urlcolor=kitgreen
]{hyperref}

\newcommand{\Eprint}[1]{\href{#1}}

\begin{document}

\title{Flavor-resolved four-fermion probes at the FCC-ee $Z$ pole}

\author{Syuhei Iguro}
\affiliation{Institute for Advanced Research, Nagoya University, Nagoya 464--8601, Japan}
\affiliation{Kobayashi-Maskawa Institute for the Origin of Particles and the Universe, Nagoya University, Nagoya 464--8602, Japan}

\begin{abstract}
The electron--positron Future Circular Collider will provide an unprecedented opportunity for precision studies at the $Z$ pole.
In addition to improving the electroweak measurements pioneered at LEP, Tera-$Z$ statistics and modern jet flavor reconstruction could make rare, flavor-resolved final states accessible as precision observables.
We study the exclusive four-body processes $e^+e^-\to Z\to Q\bar Q\ell^+\ell^-$ with $ Q=c,\,b$ and $\ell=\mu,\,\tau$ as probes of flavor-specific new physics contact interactions.
A fast analysis yields representative multi-TeV reaches for vector-current interactions involving bottom or charm quarks and muons or taus, providing a baseline for future detector-level studies.
For charm-current interactions, the projected direct sensitivities are comparable to those obtained indirectly from electroweak precision observables.
Two-dimensional fits further show that these processes retain quark-flavor information that can be poorly resolved by the renormalization-group-induced electroweak response.
These results illustrate how flavor-resolved four-body $Z$ processes can bridge the electroweak and flavor programs at FCC-ee, opening a new class of complementary precision observables.
\end{abstract}

\maketitle
Precision measurements provide an increasingly important complement to direct searches for physics beyond the Standard Model (SM).
The FCC-ee, recommended as the preferred option for CERN's next flagship collider, would substantially extend this program through unprecedented samples of Higgs, electroweak, flavor, and top-quark events \cite{EStrategy2026,FCC:2025lpp}.
At the $Z$ pole, it is expected to produce a sample approximately five orders of magnitude larger than that collected at LEP \cite{LEP:2004xhf,ALEPH:2010aa}, providing exceptional sensitivity to the virtual effects of heavy new physics.
Together with advances in tracking, vertex reconstruction, and machine-learning-based flavor tagging, these statistics also make rare and flavor-resolved final states realistic targets for precision studies \cite{Greljo:2024ytg}.

Electroweak precision observables (EWPOs) are a natural cornerstone of the FCC-ee program \cite{PDG2026}.
In the Standard Model effective field theory (SMEFT) \cite{Grzadkowski:2010es,Brivio:2017vri}, semileptonic four-fermion operators can modify these observables through renormalization-group evolution (RGE) into effective electroweak vertices.
The resulting indirect sensitivity can probe mass scales far above the collision energy \cite{DeBlas:2019qco,Allwicher:2023shc}.
It can nevertheless lose information about the flavor structure of the underlying interaction because several flavor-specific operators may generate the same effective vertex correction.
For example, $b$--$\tau$ and $c$--$\tau$ contact interactions both contribute to the effective $Z\tau\bar\tau$ coupling, and their contributions can partially cancel.
A precision measurement can strongly constrain one combination of their Wilson coefficients while leaving an orthogonal flavor direction weakly determined.
Precision on an effective vertex does not automatically imply identification of the new physics flavor structure that generated it.
Resolving this ambiguity requires observables that retain the full flavor labels of the external particles.

\begin{figure}[b] \centering 
\vspace{-0.3cm}
\includegraphics[width=0.6\columnwidth]{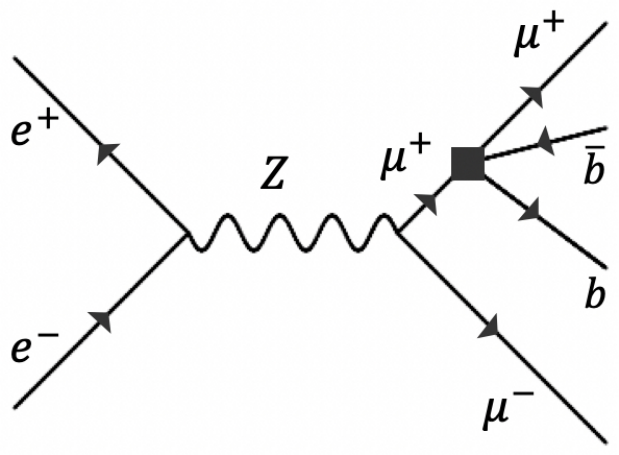} 
\vspace{-0.15cm}
\caption{ 
Representative diagram of a flavor-specific four-fermion interaction to $e^+e^-\to Z\to Q\bar Q\ell^+\ell^-$, illustrated for $Q=b$ and $\ell=\mu$.
The filled square denotes the flavor-specific four-fermion contact interaction.
}
\label{fig:diagram}
\end{figure}

To resolve this problem we propose the flavor-resolved four-body processes
\begin{align}
 e^+e^-\to Z\to Q\bar Q\ell^+\ell^-,
 \,\,\, Q=c,\,b,
 \,\,\, \ell=\mu,\,\tau,
\end{align}
as exclusive probes of flavor-specific semileptonic contact interactions at FCC-ee.
Four-fermion $Z$ decays have long been studied as rare Standard Model processes and as probes of light resonances, particularly in fully leptonic and exclusive-quarkonium final states.
Their use as inclusive, flavor-resolved precision observables for heavy semileptonic contact interactions, and as a means of resolving flavor directions left ambiguous by electroweak precision observables, remains largely unexplored.
A representative topology is illustrated in Fig.~\ref{fig:diagram}.

Unlike the EWPO sensitivity induced through RGE, these processes receive the four-fermion contribution at tree level and retain both the quark and charged-lepton flavor labels.
Their four-body kinematics also offers invariant-mass, angular, and momentum-correlation observables beyond an inclusive rate.
The huge $Z$ sample supplies the statistical sensitivity, while modern flavor reconstruction makes the underlying flavor structure experimentally accessible.
Our fast analysis finds multi-TeV sensitivity to the corresponding vector contact interactions.
For charm-current operators, the exclusive four-body reach is comparable to projected indirect EWPO sensitivities.
Realizing the projected EWPO sensitivities will require substantial advances in multi-loop calculations and refined parametric inputs for the SM predictions \cite{Chen:2020xot,Heinemeyer:2021rgq}, whereas the four-body processes probe the contact interaction directly at tree level and are subject to different theoretical systematics.
More importantly, a two-dimensional sensitivity analysis shows that the flavor-resolved processes probe directions that can remain weakly constrained by projected EWPOs.
These processes can therefore extend the FCC-ee precision program by adding flavor information and a tree-level test of the contact-interaction interpretation.\\

The flavor-resolved strategy is not restricted to a particular Lorentz structure.
As a simple demonstration, we consider the vector contact interaction
\begin{align}
{\cal L}_{\rm eff} \supset C_{VV}^{Q\ell} (\bar Q\gamma_\mu Q) (\bar\ell\gamma^\mu\ell),
\,\,\,\,[C_{VV}^{Q\ell}]={\rm TeV}^{-2}. 
\label{eq:VV_int} 
\end{align}
Purely left- and right-handed vector interactions provide additional examples and are discussed in Appendix~\ref{app:4-fermi_analysis}.
For fixed flavors $(Q,\ell)$, the interaction contributes at tree level to the corresponding exclusive process $e^+e^-\to Z\to Q\bar Q\ell^+\ell^-$.

We evaluate the parton-level rates with \textsc{MadGraph5\_aMC@NLO}~\cite{Alwall:2014hca}, using a heavy neutral vector mediator $Z^\prime$ as a simple realization of Eq.~\eqref{eq:VV_int}.
The mediator is used only to generate the contact-interaction amplitude.
Throughout the analysis, we impose the common acceptance requirement $|\eta|<2.56$ on all final-state quarks and charged leptons.
For a reference coefficient $C_0$, the cross section of the selected $e^+e^-\to Z\to Q\bar Q\ell^+\ell^-$ process can be decomposed as
\begin{align}
 \sigma(\pm C_0)=
 \sigma_{\rm SM} \pm C_0\,\sigma_{\rm int} +C_0^2\,\sigma_{\rm quad},
 \label{eq:Xs_decomposition}
\end{align}
where $\sigma_{\rm int}$ and $\sigma_{\rm quad}$ denote the coefficients of the interference and quadratic terms, respectively.
The linear EFT response is isolated through $\sigma_{\rm int}=[\sigma(+C_0)-\sigma(-C_0)]/(2C_0)$, and all quoted sensitivities are derived from this sign-odd contribution unless stated otherwise.
The contact-interaction approximation is validated to within a few percent in Appendix~\ref{app:4-fermi_analysis}, making the residual finite-mediator dependence negligible for the present sensitivity estimates.
We therefore report all results in terms of the coefficient in Eq.~\eqref{eq:VV_int}.

The dominant SM contribution contains radiation-like configurations in which an off-shell photon converts into a secondary fermion pair.
These configurations preferentially populate the soft and collinear regions of phase space, whereas a local four-fermion interaction has no corresponding low-virtuality enhancement.
We expose this difference by varying a simplified common parton-level hardness requirement, $E_{\rm cut}\equiv E_{\min}=p_{T,\min}$.
Increasing $E_{\rm cut}$ suppresses the SM rate while it enhances the interference contribution relative to the SM rate.
At still larger thresholds, the loss of accepted events dominates, producing a broad maximum in the expected sensitivity at thresholds of a few GeV.
The mild dependence on moderate variations of the angular-separation requirement further shows that the sensitivity is not generated by a finely tuned parton-level selection.
This broad sensitivity also leaves room to define control regions for validation while retaining harder regions with enhanced signal sensitivity.

\begin{figure}[t]
 \centering
 \includegraphics[width=0.75\columnwidth]{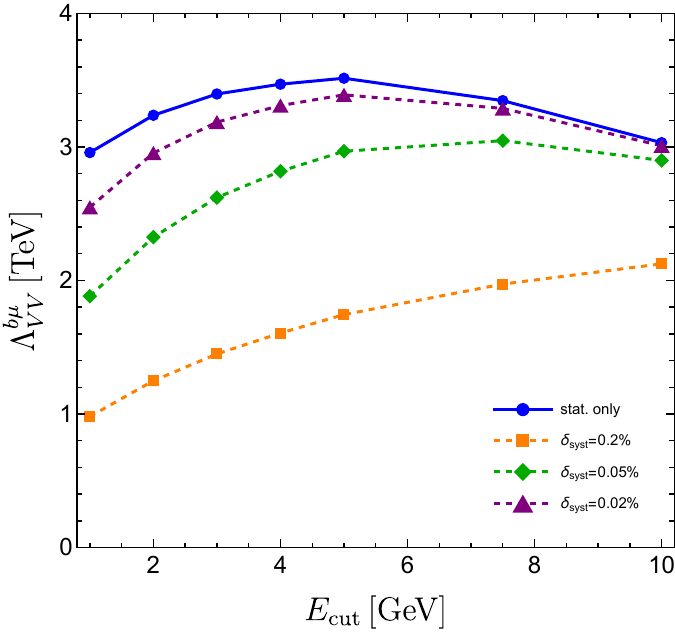}
 \vspace{-0.1cm}
 \caption{
 Illustrative $1\sigma$ sensitivity to $\Lambda_{VV}^{b\mu}$ as a function of $E_{\rm cut}$ assuming $\mathcal {L}=200~{\rm ab}^{-1}$ and unit reconstruction efficiency.
 The solid curve shows the statistics-only sensitivity obtained from the sign-odd interference contribution. 
 The orange, dark-green, and purple dashed curves additionally include benchmark systematic uncertainties of $0.2\%$, $0.05\%$, and $0.02\%$, respectively.
 \vspace{-0.1cm}
 }
 \label{fig:direct_bbmumu}
\end{figure}
For an integrated luminosity $\mathcal {L}$, the significance of the interference effect in the statistics-dominated limit is
\begin{align}
 {\cal Z}_{\rm int}(C_{VV}^{Q\ell})\simeq
 |C_{VV}^{Q\ell}| \frac{|\sigma_{\rm int}|}{\sqrt{\sigma_{\rm SM}}} \sqrt{\mathcal L\,\epsilon_{\rm rec}},
 \label{eq:int_significance}
\end{align}
where $\epsilon_{\rm rec}$ denotes an overall efficiency largely common to the SM and interference contributions.
A common efficiency loss affects the scale reach only as $\Lambda_{VV}^{Q\ell}=|C_{VV}^{Q\ell}|^{-1/2}\propto\epsilon_{\rm rec}^{1/4}$.
\begin{figure*}[t]
 \centering
 \includegraphics[width=0.47\textwidth]{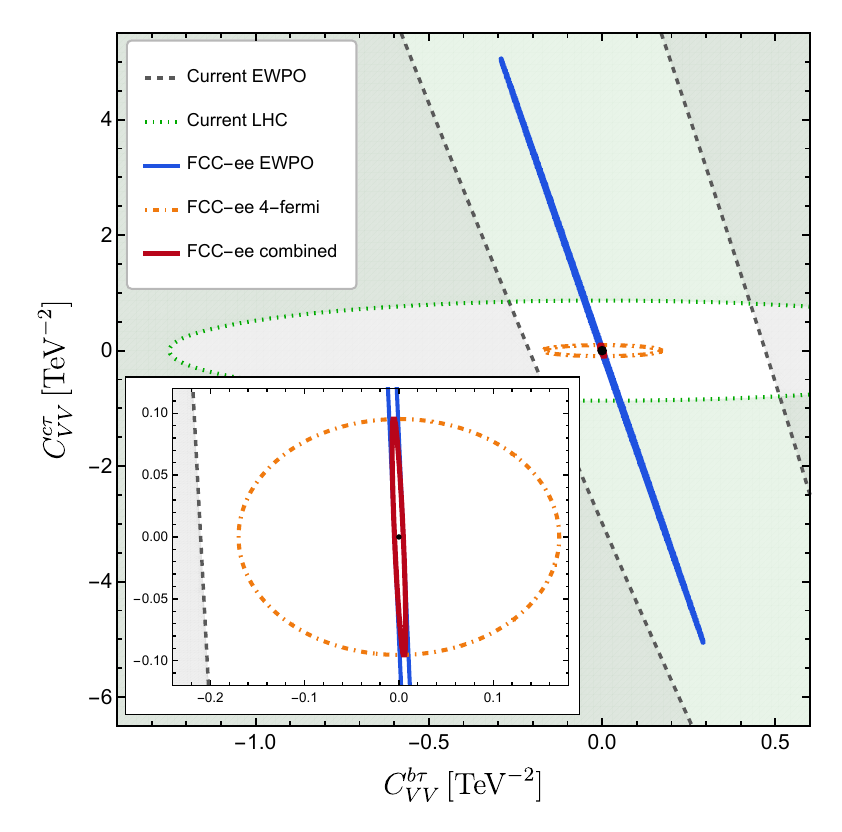} \,\,\,
 \includegraphics[width=0.47\textwidth]{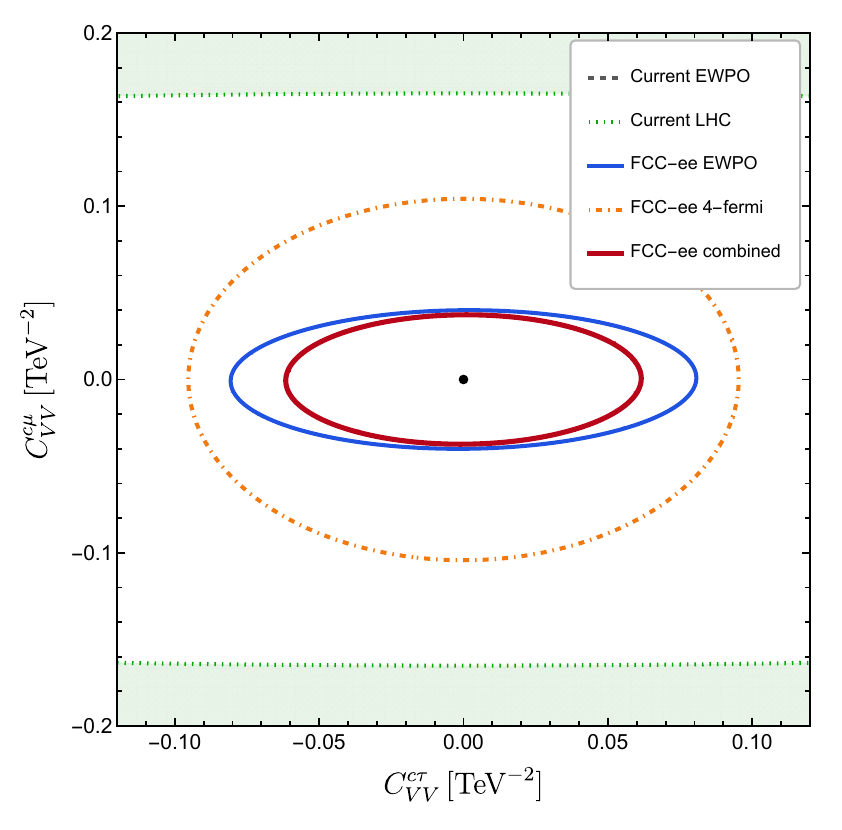}
 \caption{
 Simultaneous two-parameter statistical $95\%$ CL reaches in the $(C_{VV}^{b\tau},C_{VV}^{c\tau})$ plane (left) and the $(C_{VV}^{c\tau},C_{VV}^{c\mu})$ plane (right).
 The projected EWPO and exclusive four-body measurements probe differently oriented combinations of Wilson coefficients.
 In the left panel, the exclusive $b\bar b\tau^+\tau^-$ and $c\bar c\tau^+\tau^-$ categories probe the weak quark-flavor direction of the projected EWPO.
 In the right panel, the nominal leptonic-asymmetry inputs already separate the two lepton-flavor directions, while the exclusive processes provide an independent tree-level test.
 }
 \label{fig:2D}
\end{figure*}
For instance, an overall efficiency of $25\%$ would reduce the quoted scale reaches by only about $29\%$.
We take $\epsilon_{\rm rec}=1$ to quantify the available parton-level information, not as an assumption about detector performance.
A detector-level simulation is beyond the scope of this work.
The systematic benchmarks shown below are therefore intended to illustrate the transition from statistics- to systematics-dominated sensitivity rather than to predict the final FCC-ee experimental precision.
Even after the representative hardness requirements, the SM samples contain approximately $4\times10^6$--$2\times10^7$ events at $200~{\rm ab}^{-1}$ before reconstruction and flavor tagging.\footnote{
The corresponding LEP samples would have contained only ${\cal O}(10$--$100)$ events, and indirect constraints from EWPOs were therefore substantially stronger. 
}
In particular, flavor migration cannot be represented by a common efficiency because it mixes the bottom and charm categories and can weaken the flavor discrimination studied below.
Ongoing FCC detector studies support the feasibility of flavor-resolved bottom and charm categories~\cite{Froch:2026ftag}. 
Preliminary results indicate promising light- and heavy-flavor rejection even at signal efficiencies of $80\sim90\%$, supporting the prospect of efficient bottom and charm categorization.
Dedicated studies will nevertheless be needed to assess the tagging performance for the softer four-body topologies considered here.

Figure~\ref{fig:direct_bbmumu} illustrates the dependence of the bottom--muon sensitivity on the hardness requirement, with the underlying cross sections and finite-benchmark deviations detailed in Appendix~\ref{app:4-fermi_analysis}.
The same qualitative behavior is found in the other flavor categories.
The precise location of the maximal sensitivity is process dependent and reflects the competition between the increasing relative new physics contribution and the decreasing accepted event yield.
Nevertheless, the exclusive sensitivities remain at the multi-TeV level over a broad range of hardness requirements.

For $200~{\rm ab}^{-1}$ and unit reconstruction efficiency, the representative one-coefficient $95\%$ confidence level (CL) reaches are
\begin{align}
\left(\Lambda_{VV}^{b\mu},\Lambda_{VV}^{b\tau}, \Lambda_{VV}^{c\mu},\Lambda_{VV}^{c\tau} \right) \simeq
 \left(2.5,\,2.7,\,3.5,\,3.6\right)\,{\rm TeV},
 \label{eq:4f_1D_reaches}
\end{align}
where one coefficient is varied at a time.
The values in Eq.~\eqref{eq:4f_1D_reaches} quantify the available statistical information.
The charm channels benefit from their larger parton-level event samples, while the bottom channels expect easier flavor discrimination thanks to the large displacement from the original vertex and provide access to distinct flavor directions.

The threshold dependence suggests a strategy beyond a single optimized cut.
A high-statistics soft region can constrain the radiation-dominated SM contribution, while a harder region enhances the relative contact-interaction response.
The contact-interaction amplitude includes configurations in which the $Z$ couples to either the quark or the lepton line.
While the present analysis uses their inclusive contribution, their different kinematic structures may provide additional discriminating power in a differential analysis.
Besides, ratios of hard and soft regions within the same flavor category may reduce common normalization and tagging uncertainties while retaining sensitivity to the distinct phase-space dependence of the contact interaction.
Lepton-flavor ratios such as $\sigma(b\bar b\tau^+\tau^-)/\sigma(b\bar b\mu^+\mu^-)$ provide a complementary control by canceling the luminosity normalization and reducing correlated bottom-tagging and radiative uncertainties.
Together, the phase-space and flavor ratios would offer a path toward internally constrained precision observables rather than simple event counts.
The interference also retains information on chirality.
The chiral benchmarks show a larger response for left-handed than for right-handed currents, consistent with the chiral hierarchy of the SM $Zf\bar f$ couplings.
The operator definitions and numerical results are given in Appendix~\ref{app:4-fermi_analysis}.\\

The one-coefficient reaches do not reveal whether several flavor interactions can be identified simultaneously.
We illustrate this point in the two coefficient planes shown in Fig.~\ref{fig:2D}.
All contours correspond to simultaneous two-parameter $95\%$ confidence regions.
The current EWPO contours use present central values as implemented in \texttt{smelli}~\cite{Aebischer:2018bkb,Straub:2018kue, Aebischer:2018iyb}, while the FCC-ee projections are centered on the SM and use the projected uncertainties adopted in Ref.~\cite{Allwicher:2023shc}.
The SMEFT implementation and the construction of the current and projected EWPO likelihoods are detailed in Appendix~\ref{app:EWPO_fit}.
The four-body contours are reconstructed from the parton-level interference sensitivities by treating the flavor categories as independent Gaussian measurements.
For orientation, the panels also show representative present LHC bounds reconstructed from high-mass dilepton and ditau constraints~\cite{Allwicher:2022mcg,CMS:2021ctt,ATLAS:2025oiy,CMS:2025tlo}.
The construction of these benchmarks is detailed in Appendix~\ref{app:LHC_comparison}.

The left panel compares bottom--tau and charm--tau interactions, which generate nearly aligned corrections to the effective $Z\tau\bar\tau$ vertex.
The projected EWPO likelihood constrains predominantly the combination
\begin{equation}
 C_{VV}^{b\tau}+0.058\,C_{VV}^{c\tau},
 \label{eq:complementarity-btau-ctau-combination}
\end{equation}
and is therefore highly precise along one direction but only weakly sensitive to the approximately orthogonal cancellation direction.
The bottom- and charm-associated observables formally close the EWPO confidence region but provide only limited additional sensitivity along this quark-flavor direction.
This is a loss of flavor identification, not a loss of sensitivity to the effective $Z\tau\bar\tau$ vertex.

The exclusive processes supply the missing flavor information because the $b\bar b\tau^+\tau^-$ and $c\bar c\tau^+\tau^-$ categories retain the quark-flavor labels.
Their standalone reaches are weaker than the best-constrained EWPO direction, whose projected sensitivity is conditional on corresponding progress in precision SM calculations.
However, the four-body information is aligned close to the flavor axes and therefore directly constrains the EWPO cancellation direction.
The indirect sensitivity is particularly strong in the bottom--tau direction because operators involving the third-generation quark doublet receive top-Yukawa-enhanced running into electroweak vertices.

Combining the flavor-resolved measurements with the projected FCC-ee EWPO likelihood, while allowing both Wilson coefficients to vary, gives the simultaneous two-parameter $95\%$ CL reaches
\begin{equation}
 \Lambda_{VV}^{b\tau}>12~{\rm TeV},
 \,\,\,\, \Lambda_{VV}^{c\tau}>3.2~{\rm TeV}.
 \label{eq:complementarity-bctau}
\end{equation}
The bottom--tau reach is driven by the EWPO precision and becomes individually meaningful once the four-body measurements constrain the quark-flavor cancellation direction.

The right panel shows the sensitivity to charm--tau and charm--muon interactions.
Here the quark flavor is fixed, while the projected $A_\tau$ and $A_\mu$ information probes two approximately independent charged-lepton directions.
The nominal projected EWPO likelihood is consequently well conditioned and gives simultaneous marginalized reaches of $\Lambda_{VV}^{c\tau}>3.5~{\rm TeV}$ and $\Lambda_{VV}^{c\mu}>5.0~{\rm TeV}$.
The exclusive charm--tau reach is numerically comparable.
The exclusive modes nevertheless test the same flavor interactions through tree-level four-fermion amplitudes and do not rely on the same pseudo-observable interpretation.
At an unpolarized collider, the projected $A_\mu$ input must be interpreted together with the relation $A_{\rm FB}^{0,\mu}=3A_eA_\mu/4$.
A complete FCC-ee analysis will therefore require the covariance among the leptonic pseudo-observables.\footnote{
As a robustness test, removing the separately included $A_{\rm FB}^{0,\mu}$ input while retaining the projected $A_\mu$ information has a negligible impact on the $C_{VV}^{c\mu}$ projection.
The nominal $5~{\rm TeV}$ reach is thus not driven by the separately included $A_{\rm FB}^{0,\mu}$ term, although its final experimental interpretation remains covariance dependent.}

The two planes illustrate the complementary roles of precision and flavor resolution. 
EWPOs constrain effective electroweak vertices with exceptional precision but may retain poorly constrained flavor directions. 
The exclusive four-body categories preserve the external-state flavor labels and provide complementary tests of the underlying interactions.\\

The present results provide a parton-level statistical baseline for the sensitivity of flavor-resolved four-body categories.
A realistic analysis must account for higher-order QED, electroweak, and QCD radiation, hadronization, flavor-dependent tagging efficiencies and migration, tau reconstruction, and reducible backgrounds.
Recent NNLO calculations of four-jet production in $e^+e^-$ annihilation illustrate the progress toward precision predictions for complex multi-particle final states~\cite{Chen:2026jxf}.
Extending such efforts to $e^+e^-\to Z\to Q\bar Q\ell^+\ell^-$ will be important for exploiting the statistical precision of the Tera-$Z$ sample.
These effects will modify the numerical reach, but the broad range of hardness requirements with appreciable sensitivity and the weak fourth-root dependence on the overall efficiency suggest that the underlying physics opportunity is not tied to a finely tuned selection.
A detector-level multi-bin analysis combining absolute rates, hard-to-soft ratios, and lepton-flavor ratios is therefore a natural next step \cite{WIP}.
The rate analysis considered here is only the first layer of the available information.
Angular and polarization observables would improve the signal--background discrimination.
Furthermore, four-body final states allow CP-odd momentum correlations that can probe CP-violating Wilson coefficients, which are not accessible through conventional CP-even two-body observables.
Improved multiclass tagging may extend the same strategy to strange-quark currents. 
Related exclusive final states could also probe four-lepton interactions or interactions involving invisible and semi-visible particles, providing access to dark-sector scenarios such as those considered in Refs.~\cite{Okawa:2020jea,Iguro:2022tmr,Higuchi:2023kbt}.

Flavor-resolved four-body $Z$ processes can thus broaden the FCC-ee precision program through qualitatively new observables. 
They preserve information about the underlying flavor structure and probe the interactions at tree level.
Their multi-dimensional kinematics also offers internal strategies for controlling systematic uncertainties. 
The Tera-$Z$ sample can therefore provide not only more precise versions of LEP measurements, but also qualitatively new tests of heavy new physics.

\section*{Acknowledgements}
S.\,I. thanks Xunwu Zuo, Teppei Kitahara, Tim Kretz, Florian Kretz, Hantian Zhang, and Hiroyasu Yonaha for inspiring discussions.
S.\,I. also appreciates the workshop ``{\it{\href{https://indico.cern.ch/event/1644557/timetable/}{Flavours at FCC-ee}}}'' which motivated this work.
This work is supported by JSPS KAKENHI Grant Numbers 22K21347, 24K23939 and 25K17385 and the Toyoaki Scholarship Foundation.

\appendix
\section{Analysis of 4-fermion processes} 
\label{app:4-fermi_analysis} 
\subsection{Mediator implementation and EFT validation}
\label{app:mediator_validation}
The parton-level analysis is performed at the $Z$ pole with an integrated luminosity of ${\cal L}=200~{\rm ab}^{-1}$.
A phenomenological heavy neutral vector boson is used as a numerical generator of the four-fermion amplitude.
In the heavy-mass limit, $m_{Z^\prime}^2\gg s$, its exchange produces
\begin{align}
 {\cal L}_{\rm eff} \supset
 C_{VV}^{Q\ell} (\bar Q\gamma_\mu Q) (\bar\ell\gamma^\mu\ell),
 \,\,\,\, C_{VV}^{Q\ell} = \frac{g_Qg_\ell}{M_{Z'}^2},
\end{align}
up to the overall sign convention of the mediator couplings.
The mediator is used only as a simulation tool and is not interpreted as an on-shell signal.

For a reference coefficient $C_0$, the cross section is decomposed as
in Eq.~\eqref{eq:Xs_decomposition}.
Samples with opposite coefficient signs determine the interference
coefficient through
\begin{align}
 \sigma_{\rm int}=
 \frac{\sigma(+C_0)-\sigma(-C_0)}{2C_0},
 \label{eq:app_interference}
\end{align}
while the quadratic coefficient is obtained from
\begin{align}
 \sigma_{\rm quad} =
 \frac{ \sigma(+C_0)+\sigma(-C_0)-2\sigma_{\rm SM} }{2C_0^2}.
 \label{eq:app_quadratic}
\end{align}
The quoted EFT sensitivities in Eq.~\eqref{eq:4f_1D_reaches} are derived from $\sigma_{\rm int}$ and therefore retain only the contribution linear in the dimension-six coefficient.
The finite quadratic contribution in the reference samples is not included in the asymptotic scale reach.

The validity of the contact-interaction approximation is tested in the $b\bar b\mu^+\mu^-$ channel with
\begin{align}
 E_{\min}=p_{T,\min}=3~{\rm GeV},
 \,\,\,\,
 \Delta R_{\min}=0.2.
 \label{eq:app_validation_selection}
\end{align}
For mediator masses $M_{Z'}=300$, $424$, and $600~{\rm GeV}$, we confirmed the expected $\sigma_{\rm int}\propto M_{Z'}^{-2}$ scaling within $3\%$.
At the reference point $M_{Z'}=300~{\rm GeV}$, the quadratic contribution is sub-leading to the interference for the representative selection considered here.
Its relative importance decreases further at the multi-TeV scales.
All final results are expressed in terms of the coefficients $C_{VV}^{Q\ell}$ defined in Eq.~\eqref{eq:VV_int}.

\subsection{Kinematic dependence and cross sections}
\label{app:reference_Xs}
Having validated the contact-interaction approximation, we next examine the kinematic dependence of the SM and interference contributions.
The dominant SM contribution contains radiation-like configurations in which a virtual photon converts into a secondary fermion pair.
These configurations preferentially populate regions with low final-state energies, transverse momenta, and pairwise virtualities.
The local contact interaction has no corresponding low-virtuality enhancement and therefore exhibits a different dependence on the four-body kinematics.

To expose this difference, we impose the common cumulative hardness requirement
\begin{align}
 E_{\rm cut} = 1,\,2,\,3,\,4,\,5,\,7.5,\,10~{\rm GeV},
 \label{eq:app_hardness_scan}
\end{align}
together with the reference angular-separation requirement $\Delta R_{\min}=0.2$.
For simplicity, the energy and transverse-momentum thresholds are varied simultaneously.
The scan should therefore be interpreted as a variation of a common parton-level hardness requirement rather than of a single kinematic variable.
In the \textsc{MadGraph5} simulation, $b$ and $\tau$ are treated as massive, while $c$ and $\mu$ are treated as massless.

The SM cross section decreases rapidly as $E_{\rm cut}$ is increased, while the interference contribution generally becomes larger relative to the SM rate. 
At sufficiently large thresholds, however, the reduction in the accepted event yield dominates. 
The competition between these effects produces broad sensitivity maxima at thresholds of a few GeV. 
Table~\ref{tab:Xs} illustrates this tradeoff by comparing a loose selection with the harder selection that maximizes the statistics-only interference sensitivity within the scan for each channel.
For $M_{Z'}=300~{\rm GeV}$ and unit mediator couplings, the reference coefficient is
\begin{align}
 C_0 =
 \frac{1}{M_{Z'}^2} \simeq 11.1~{\rm TeV}^{-2}.
 \label{eq:app_coefficient}
\end{align}
The total fractional deviations of the finite benchmark samples are defined by
\begin{align}
 \delta_\pm=
 \frac{\sigma(\pm C_0)-\sigma_{\rm SM}} {\sigma_{\rm SM}}.
 \label{eq:app_fractional_deviation}
\end{align}

\begin{table*}[t] \centering
\renewcommand{\arraystretch}{1.12} 
\begin{tabular}{c|c|ccc|cc} \hline\hline
\,channel\, & \,$E_{\rm cut}$ [GeV]\, & \,$\sigma_{\rm SM}$ [pb]\, & \,$\sigma_+$ [pb]\, & \,$\sigma_-$ [pb]\, & \,$\delta_+$ [\%]\, & \,$\delta_-$ [\%]\, \\ \hline
$b\bar b\mu^+\mu^-$ & $1$ & $0.102$ & $0.0999$ & $0.104$ & $-2.05$ & $+2.25$ \\
& $5$ & $0.0194$ & $0.0181$ & $0.0208$ & $-6.41$ & $+7.55$ \\ \hline 
$b\bar b\tau^+\tau^-$ & $1$ & $0.0327$ & $0.0308$ & $0.0349$ & $-5.76$ & $+6.56$ \\
& $3$ & $0.0223$ & $0.0208$ & $0.0242$ & $-7.03$ & $+8.15$ \\ \hline
$c\bar c\mu^+\mu^-$ & $1$ & $0.553$ & $0.563$ & $0.543$ & $+1.86$ & $-1.74$ \\
& $5$ & $0.0978$ & $0.104$ & $0.0923$ & $+6.17$ & $-5.64$ \\ \hline 
$c\bar c\tau^+\tau^-$ & $1$ & $0.344$ & $0.354$ & $0.335$ & $+2.91$ & $-2.59$ \\ & $4$ & $0.111$ & $0.118$ & $0.105$ & $+6.29$ & $-5.84$ \\ \hline\hline
\end{tabular} 
\caption{ 
SM cross sections and fractional deviations for the finite $M_{Z'}=300~{\rm GeV}$ benchmark with unit mediator couplings.
For each channel, the first row uses $E_{\rm cut}=1~{\rm GeV}$, while the second uses the threshold that maximizes the statistics-only interference sensitivity within the scan.
The subscripts $+$ and $-$ denote the two signs of $C_{VV}^{Q\ell}$.
The harder selection reduces the SM rate while enhancing the relative contact-interaction response.
} 
\label{tab:Xs}
\end{table*}

At $E_{\rm cut}=1~{\rm GeV}$, the SM cross sections range from approximately $33~{\rm fb}$ in the $b$--$\tau$ channel to $553~{\rm fb}$ in the $c$--$\mu$ channel.
At $200~{\rm ab}^{-1}$, these rates correspond to approximately $6.5\times10^6$ to $1.1\times10^8$ events before reconstruction and flavor tagging.
The finite benchmark deviations are approximately $2$--$7\%$ in this loose region.

The harder selections reduce the SM cross sections to approximately $19$--$111~{\rm fb}$, corresponding to approximately $4\times10^6$ to $2\times10^7$ parton-level events at $200~{\rm ab}^{-1}$.
At the same time, the magnitudes of the finite benchmark deviations increase to approximately $6$--$8\%$.
The enhancement is particularly visible in the muon channels.
The magnitude of the shift increases from approximately $2\%$ to $7\%$ for the $b$--$\mu$ channel and from approximately $2\%$ to $6\%$ for the $c$--$\mu$ channel.

This behavior reflects the preferential removal of soft and collinear SM configurations, including photon-conversion topologies, to which a local four-fermion interaction has no analogous low-virtuality enhancement.
The comparison is intended to expose this kinematic tradeoff rather than to construct a unique optimized threshold.
The loose region retains a large event sample and may help constrain the radiation-dominated SM contribution, while the harder region enhances the relative contact-interaction response.
A realistic analysis should therefore combine several exclusive kinematic regions rather than rely on a single cumulative event rate.

The small asymmetry between $|\delta_+|$ and $|\delta_-|$ arises from the contribution quadratic in the finite benchmark coefficient.
The quadratic fraction, 
\begin{align}
\delta_{\rm quad} = \frac{ \sigma(+C_0)+\sigma(-C_0)-2\sigma_{\rm SM} }{2\sigma_{\rm SM}},
\end{align}
remains below $0.8\%$ for the representative entries in Table~\ref{tab:Xs} and is therefore sub-leading in the finite benchmark samples.
The sensitivities quoted in the main text are obtained from the sign-odd combination in Eq.~\eqref{eq:app_interference}, which removes the quadratic term exactly within the generated samples.

Moderate variations of the angular-separation requirement do not drastically change the threshold dependence.
This stability indicates that the observed sensitivity is not generated by a finely tuned parton-level separation cut.
The separation requirement should nevertheless be replaced by an infrared-safe object and jet definition in a higher-order and detector-level analysis.

\begin{figure*}[t]
 \centering
 \begin{tabular}{ccc}
  \includegraphics[width=0.31\textwidth]{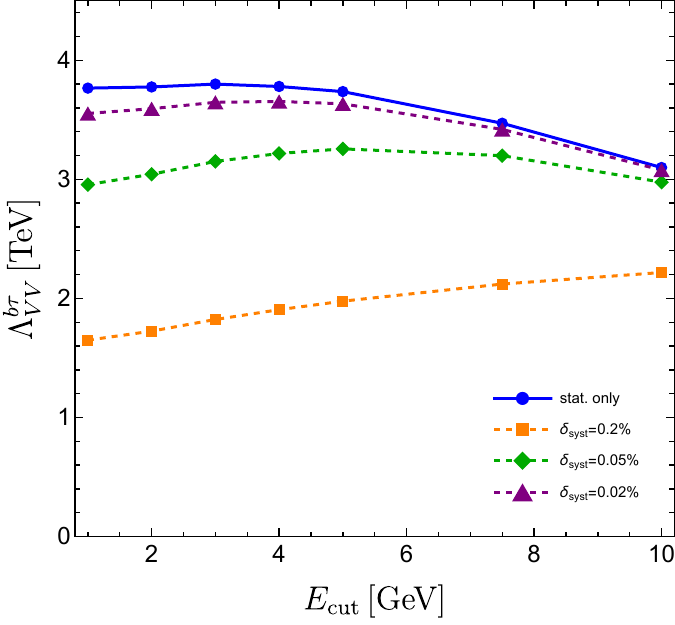}\,\,\,
  \includegraphics[width=0.31\textwidth]{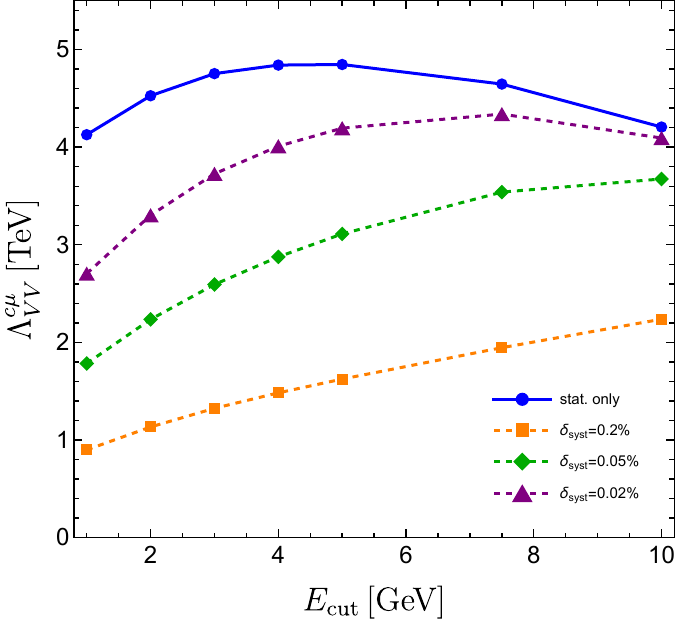}\,\,\,
  \includegraphics[width=0.31\textwidth]{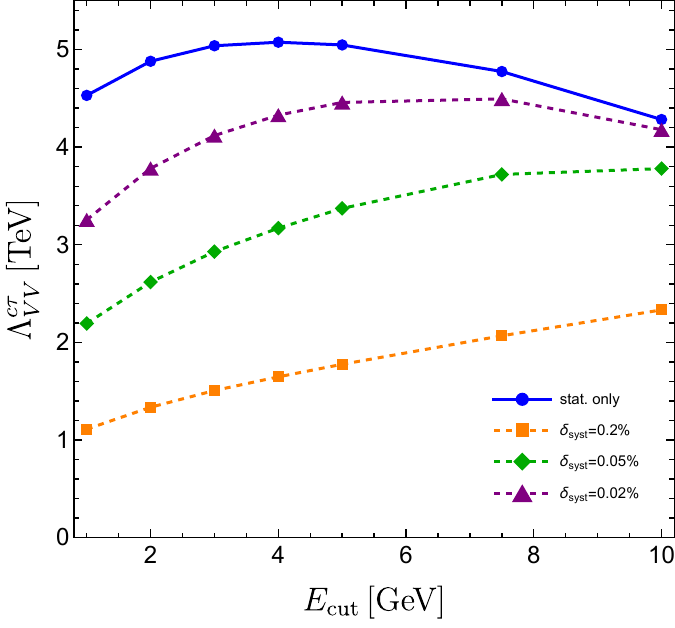}
 \end{tabular}
 \caption{
 Projected sensitivities in the $b\bar b\tau^+\tau^-$, $c\bar c\mu^+\mu^-$, and $c\bar c\tau^+\tau^-$ channels from left to right.
 The assumptions and curve conventions are the same as in Fig.~\ref{fig:direct_bbmumu}.
 All three channels exhibit broad sensitivity maxima at thresholds of a few GeV.
 }
 \label{fig:direct_QQll}
\end{figure*}

The threshold dependence of the remaining semileptonic channels is shown in Fig.~\ref{fig:direct_QQll}.
Together with the $b$--$\mu$ example in Fig.~\ref{fig:direct_bbmumu}, the four channels exhibit the same broad pattern, although the location and height of the sensitivity maximum are process dependent.

The common hardness requirement suppresses the radiation-dominated SM contribution while enhancing the relative contact-interaction response.
At larger thresholds, this gain is offset by the decreasing accepted event yield.
The persistence of multi-TeV sensitivity over a broad range of thresholds motivates the statistical treatment developed in the next subsection.

We now translate the interference responses discussed above into one-parameter sensitivity estimates.
For a coefficient $|C_{VV}^{Q\ell}|=\left(\Lambda_{VV}^{Q\ell}\right)^{-2}$, the significance in the statistics-dominated limit is given in Eq.~\eqref{eq:int_significance}.
Equivalently, the reference $n$-standard-deviation scale reach is
\begin{align}
 \Lambda_{VV}^{n\sigma} =
 \left[ 
 \frac{|\sigma_{\rm int}|\,\sqrt{{\cal L}\,\epsilon_{\rm rec}}} {n\,\sqrt{\sigma_{\rm SM}}}
 \right]^{1/2},
 \label{eq:app_scale_reach}
\end{align}
where $\sigma_{\rm int}$ is the coefficient of the term linear in $C_{VV}^{Q\ell}$, as defined in Eq.~\eqref{eq:app_interference}.
The dimensions of $\sigma_{\rm int}$ are therefore those of a cross section multiplied by ${\rm TeV}^{2}$.

Using $n=1.96$ for a two-sided one-parameter $95\%$ confidence criterion, ${\cal L}=200~{\rm ab}^{-1}$, and
$\epsilon_{\rm rec}=1$, we obtain
\begin{align}
 \left( \Lambda_{VV}^{b\mu},\, \Lambda_{VV}^{b\tau},\, \Lambda_{VV}^{c\mu},\, \Lambda_{VV}^{c\tau} \right) =
 \left( 2.51,\,2.71,\,3.46,\,3.62 \right)~{\rm TeV}.
 \label{eq:app_direct_reaches}
\end{align}
The representative thresholds used for these values are $E_{\rm cut}=3\sim5~{\rm GeV}$ depending on channels.
The full threshold dependence is shown in Figs.~\ref{fig:direct_bbmumu} and \ref{fig:direct_QQll}.
The charm channels benefit from their larger accepted parton-level samples, while the bottom channels provide sensitivity to distinct quark-flavor directions.

For an efficiency common to the SM and interference contributions, Eq.~\eqref{eq:app_scale_reach} implies
\begin{align}
 \Lambda_{VV}^{Q\ell}(\epsilon_{\rm rec}) = \Lambda_{VV}^{Q\ell}(1)\, \epsilon_{\rm rec}^{1/4}.
 \label{eq:app_efficiency_scaling}
\end{align}
For example, an overall efficiency of $25\%$ reduces the scale reach by the factor $0.25^{1/4}\simeq0.71$, corresponding to a reduction of approximately $29\%$.
This rescaling applies only when the efficiency acts approximately as a common multiplicative factor.
It does not account for flavor migration, channel-dependent acceptance, or reducible backgrounds, which require a more dedicated analysis.
The large samples of two-body $Z\to b\bar b$ and $Z\to c\bar c$ decays may provide data-driven control samples for calibrating the flavor-tagging efficiencies and migration probabilities.
Dedicated studies are nevertheless required to validate the transfer of these calibrations to the softer and more complex four-body topologies considered here.

To illustrate the impact of a systematic uncertainty, we define the total fractional uncertainty on the SM rate as
\begin{align}
 \epsilon_{\rm tot}^2 = \frac{1}{{\cal L}\,\epsilon_{\rm rec}\,\sigma_{\rm SM}} + \delta_{\rm syst}^2,
\end{align}
where $\delta_{\rm syst}$ denotes an additional systematic uncertainty.
The corresponding scale reach is
\begin{align}
 \Lambda_{VV}^{n\sigma}  = 
 \left[ \frac{|\sigma_{\rm int}|} {n\,\sigma_{\rm SM}\,\epsilon_{\rm tot}} \right]^{1/2}.
\end{align}
The dashed curves show representative choices of $\delta_{\rm syst}=0.02\%$, $0.05\%$, and $0.2\%$.
These values should be regarded as diagnostic benchmarks rather than projections of the achievable experimental precision
The first two are comparable to the ${\cal O}(0.02\%$--$0.05\%)$ statistical precision of the selected parton-level samples, while $0.2\%$ illustrates a clearly systematics-dominated regime.
The values in Eq.~\eqref{eq:app_direct_reaches} therefore quantify the statistical information available in the idealized parton-level flavor categories.

\subsection{Chiral benchmark interactions}
\label{app:chiral_benchmarks}
The flavor-resolved rate analysis is not restricted to the vector interaction considered in the main text.
To illustrate its sensitivity to chirality, we define
\begin{align}
 {\cal O}_{XY}^{Q\ell} = (\bar Q\gamma_\mu P_X Q)(\bar\ell\gamma^\mu P_Y\ell),  \,\,\,\,\, X,Y=L,R,
 \label{eq:app_chiral_int}
\end{align}
where $P_{L,R}=(1\mp\gamma_5)/2$.
The $LL$ and $RR$ interactions correspond to the $(V-A)\times(V-A)$ and $(V+A)\times(V+A)$ structures, respectively.
The vector-current interaction studied in the main text is the correlated combination
\begin{align}
 {\cal O}_{VV}^{Q\ell} = {\cal O}_{LL}^{Q\ell} +{\cal O}_{LR}^{Q\ell} +{\cal O}_{RL}^{Q\ell} +{\cal O}_{RR}^{Q\ell}.
\end{align}
Consequently, the $VV$, $LL$, and $RR$ scale reaches quoted below refer to different operator normalizations and should be compared only after accounting for the chiral projectors in Eq.~\eqref{eq:app_chiral_int}.

Although the dominant photon-conversion contribution is vector-like, the full SM amplitude and its interference with the contact interaction remain sensitive to chirality.
The primary fermion pair is produced through the chiral SM $Zf\bar f$ couplings, and the contact amplitude can attach to either the quark or lepton line.
The larger left-handed $Zf\bar f$ couplings therefore produce a larger $LL$ than $RR$ response, particularly for bottom currents, for which the right-handed SM coupling is small.

The representative one-parameter scale reaches inferred from the available samples are
\begin{align}
\begin{array}{c|ccc}
 \text{channel} & \Lambda_{VV}^{95\%}\,[{\rm TeV}] & \Lambda_{LL}^{95\%}\,[{\rm TeV}] & \Lambda_{RR}^{95\%}\,[{\rm TeV}] \\ \hline
 b\bar b\mu^+\mu^-   & 2.51 & 1.25 & 0.45 \\
 b\bar b\tau^+\tau^- & 2.71 & 1.60 & 0.74 \\
 c\bar c\mu^+\mu^-   & 3.46 & 1.77 & 1.11 \\
 c\bar c\tau^+\tau^- & 3.62 & 1.94 & 1.20
\end{array}
\label{eq:app_semilep_chiral_reaches}
\end{align}
The hierarchy $\Lambda_{LL}>\Lambda_{RR}$ is common to all four semileptonic channels and is most pronounced for bottom currents.

The chiral entries should not be placed on the same quantitative footing as the $VV$ results.
For the $VV$ direction, samples with both coefficient signs isolate the interference according to Eq.~\eqref{eq:app_interference}.
For most $LL$ and $RR$ directions, only one coefficient sign is simulated, and the working estimate assumes interference dominance,
\begin{align}
 \sigma_{\rm int}^{X} \simeq \frac{\sigma_X-\sigma_{\rm SM}}{C_0}, \,\,\,\, X=LL,\,\,RR.
\end{align}
The scale reaches in the $LL$ and $RR$ columns of Eq.~\eqref{eq:app_semilep_chiral_reaches} should therefore be regarded as indicative.
They are not used in the principal two-dimensional flavor analysis.
No axial--axial reach is quoted because the generated $10^6$ samples do not distinguish a suppressed interference contribution from the quadratic term.

\subsection{Four-lepton final state}
\label{app:mumutautau}
The same analysis strategy can be applied to the four-lepton interaction
\begin{align}
 {\cal L}_{\rm eff}\supset
 C_{VV}^{\mu\tau} (\bar\mu\gamma_\alpha\mu)(\bar\tau\gamma^\alpha\tau).
\end{align}
This channel does not require heavy-quark flavor tagging and therefore provides a complementary test with experimental systematics different from those of the semileptonic modes.

\begin{figure}[t]
 \centering
 \includegraphics[width=0.78\columnwidth]{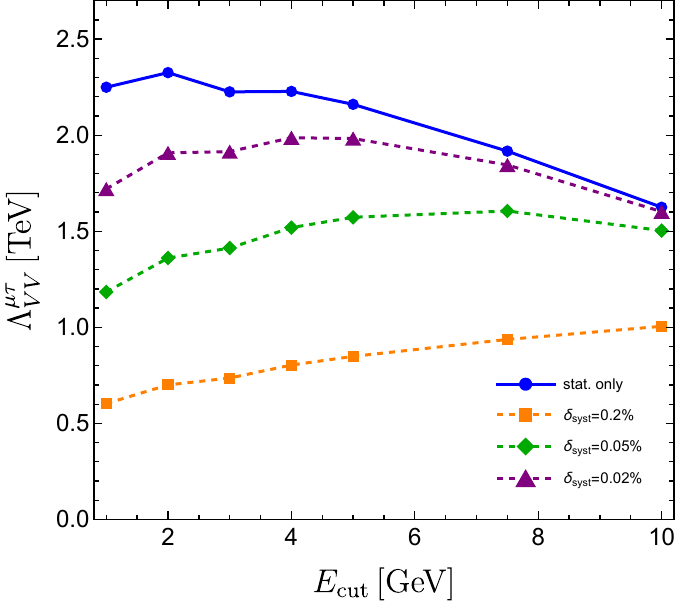}
 \caption{
 Projected interference-only sensitivity to $\Lambda_{VV}^{\mu\tau}=|C_{VV}^{\mu\tau}|^{-1/2}$ as a function of $E_{\rm cut}$, assuming ${\cal L}=200~{\rm ab}^{-1}$ and unit reconstruction efficiency.
 The solid curve shows the statistics-only sensitivity obtained from the sign-odd interference contribution.
 }
 \label{fig:app_mumutau_threshold_scan}
\end{figure}

Samples with opposite coefficient signs are used to isolate the
interference according to Eq.~\eqref{eq:app_interference}.
The resulting one-parameter $95\%$ CL scale reaches are
\begin{align}
\begin{array}{c|ccccccc}
 E_{\rm cut}\,[{\rm GeV}] & 1 & 2 & 3 & 4 & 5 & 7.5 & 10 \\ \hline
 \Lambda_{\mu\tau}^{95\%}\,[{\rm TeV}] & 1.61 & 1.66 & 1.59 & 1.59 & 1.54 & 1.37 & 1.16
\end{array}
\end{align}
The best sensitivity is obtained near $E_{\rm cut}=2~{\rm GeV}$.
The full threshold dependence is shown in Fig.~\ref{fig:app_mumutau_threshold_scan}.
Compared with the semileptonic channels, the sensitivity has a softer and milder threshold dependence, reflecting the different balance of radiation topologies and available phase space in the purely leptonic final state.

For comparison, the current EWPO sensitivity to the correlated $VV$ direction is approximately
\begin{align}
 \Lambda_{\mu\tau,\,{\rm current\ EWPO}}^{95\%} \simeq 0.32~{\rm TeV},
\end{align}
while the nominal experimental-error-only FCC-ee projection gives
\begin{align}
 \Lambda_{\mu\tau,\,{\rm future\ EWPO}}^{95\%} \simeq 4.81~{\rm TeV}.
\end{align}
The projected EWPO sensitivity is driven primarily by the assumed future $A_\mu$ and $A_\tau$ information and is subject to the covariance and SM theory qualifications discussed in Appendix~\ref{app:EWPO_fit}.
The four-lepton process does not exceed this nominal indirect reach, but it probes the interaction through a tree-level four-body amplitude. 
It therefore provides an independent exclusive test. 
Direct LHC sensitivity to $C_{VV}^{\mu\tau}$ is expected to remain limited by the small leptonic parton luminosity in proton--proton collisions.

\section{Electroweak fits}
\label{app:EWPO_fit} 

\subsection{SMEFT mapping and implementation}
\label{app:SMEFT_mapping}
The direct four-body analysis is formulated in terms of the broken-phase coefficients $C_{VV}^{Q\ell}$ defined in Eq.~\eqref{eq:VV_int}.
For the electroweak interpretation, these interactions are embedded into correlated Warsaw-basis directions~\cite{Grzadkowski:2010es}.
We use dimensionful Wilson coefficients with the convention
\begin{align}
 {\cal L}_{\rm SMEFT} \supset \sum_a C_a {\cal O}_a,
 \,\,\,\, [C_a]={\rm TeV}^{-2}.
 \label{eq:app-smeft-coefficient-convention}
\end{align}
The coefficients are fixed at the ultraviolet input scale $\mu_{\rm UV}=3~{\rm TeV}$.
For a coefficient $C_{VV}^{Q\ell}$, we define the shorthand scale $\Lambda_{VV}^{Q\ell}=|C_{VV}^{Q\ell}|^{-1/2}$.
For a charged-lepton generation $\alpha$ and a down-type quark generation $i$, the vector-current direction is defined by
\begin{align}
 [C_{\ell q}^{(1)}]_{\alpha\alpha ii}&=
 [C_{qe}]_{ii\alpha\alpha} = [C_{\ell d}]_{\alpha\alpha ii} = [C_{ed}]_{\alpha\alpha ii} = C_{VV}^{d_i\ell_\alpha}, \nonumber\\
 [C_{\ell q}^{(3)}]_{\alpha\alpha ii}  &=0.
 \label{eq:app_down_type_VV}
\end{align}
For a charged-lepton generation $\alpha$ and an up-type quark generation $i$, the corresponding direction is defined by
\begin{align}
 [C_{\ell q}^{(1)}]_{\alpha\alpha ii} &=
 [C_{qe}]_{ii\alpha\alpha} = [C_{\ell u}]_{\alpha\alpha ii} = [C_{eu}]_{\alpha\alpha ii} = C_{VV}^{u_i\ell_\alpha}, \nonumber\\
 [C_{\ell q}^{(3)}]_{\alpha\alpha ii} &=0.
 \label{eq:app_up_type_VV}
\end{align}
These equalities reproduce the charged-lepton component $(\bar Q\gamma_\mu Q)(\bar\ell\gamma^\mu\ell)$ by assigning the same coefficient to its $LL$, $LR$, $RL$, and $RR$ components.

The four semileptonic benchmark directions are obtained from $(i,\alpha)=(3,3)$ for $b\tau$, $(3,2)$ for $b\mu$, $(2,3)$ for $c\tau$, and $(2,2)$ for $c\mu$.
All Wilson coefficients not listed in Eqs.~\eqref{eq:app_down_type_VV} and \eqref{eq:app_up_type_VV} are set to zero at the input scale.

The Wilson coefficients are evolved from $3~{\rm TeV}$ to the electroweak scale with \texttt{wilson}~2.5.2~\cite{Aebischer:2018bkb}.
The evolution uses the numerical solution of the SMEFT renormalization-group equations with
\begin{align}
 \texttt{smeft\_accuracy}=``\texttt{integrate}".
\end{align}
Electroweak observables are evaluated with \texttt{flavio}~2.7.1~\cite{Straub:2018kue}.
The present electroweak likelihood is evaluated with \texttt{smelli}~2.4.3~\cite{Aebischer:2018iyb}.
The projected FCC-ee likelihood is constructed separately from the linearized observable responses and the future pseudo-observable uncertainties described in Appendix~\ref{app:EWPO_matrices}.

\subsection{Current and projected electroweak likelihoods}
\label{app:EWPO_matrices}
For Wilson coefficients collected in the vector $\boldsymbol C$, the electroweak likelihood is represented locally by the quadratic form 
\begin{align}
\Delta\chi^2 = 
(\boldsymbol C-\widehat{\boldsymbol C})^T F (\boldsymbol C-\widehat{\boldsymbol C}), 
\end{align}
where $\widehat{\boldsymbol C}$ denotes the local best-fit point and $F$ is the symmetric matrix describing the local curvature of the likelihood.

The current EWPO likelihood is obtained from the present electroweak data implemented in \texttt{smelli} and retains the measured central values.
Its best-fit point $\widehat{\boldsymbol C}$ therefore need not coincide with the SM point.
The projected FCC-ee likelihood is instead an Asimov sensitivity centered on the SM.
For the $(C_{VV}^{b\tau},C_{VV}^{c\tau})$ plane, the nominal observable set consists of $R_\tau$, $A_\tau$, $A_{\rm FB}^{0,\tau}$, $R_b$, $A_{\rm FB}^{0,b}$, $R_c$, and $A_{\rm FB}^{0,c}$.
The relative uncertainties are taken from Table~9 of Ref.~\cite{Allwicher:2023shc}. 
Thus, bottom- and charm-associated observables are included, although their projected precision in this benchmark is insufficient to remove the strong correlation generated by the dominant tau-sector response.
For the $(C_{VV}^{c\tau},C_{VV}^{c\mu})$ plane, the nominal likelihood additionally uses the corresponding muon pseudo-observables, including the effective projected $A_\mu$ input.
The current and projected likelihoods represent alternative electroweak scenarios and are not added to one another.

For completeness, we give the local quadratic expansions of the current likelihood relative to the SM point.
In the $(C_{VV}^{b\tau},C_{VV}^{c\tau})$ plane, this expansion is
\begin{align}
 \chi^2_{\rm current}(C_{VV}^{b\tau}&,C_{VV}^{c\tau})
 -\chi^2_{\rm current}(0,0)=\nonumber\\
 & -10.1\,C_{VV}^{b\tau}
 -0.830\,C_{VV}^{c\tau}
 \nonumber\\
 &+38.6\,(C_{VV}^{b\tau})^2 +4.71\,C_{VV}^{b\tau}C_{VV}^{c\tau}
 \nonumber\\
 &+ 0.151\,(C_{VV}^{c\tau})^2.
 \label{eq:app-current-btau-ctau-ewpo}
\end{align}
In the $(C_{VV}^{c\tau},\,C_{VV}^{c\mu})$ plane, the current local expansion is
\begin{align}
 \chi^2_{\rm current}(C_{VV}^{c\tau}&,\,C_{VV}^{c\mu})
 -\chi^2_{\rm current}(0,0)=\nonumber\\
 &-0.710\,C_{VV}^{c\tau} -0.492\,C_{VV}^{c\mu}
 \nonumber\\
 &+ 0.167\,(C_{VV}^{c\tau})^2 +0.0699\,C_{VV}^{c\tau}C_{VV}^{c\mu}
 \nonumber\\
 &+ 0.0676\,(C_{VV}^{c\mu})^2.
 \label{eq:app-current-ctau-cmu-ewpo}
\end{align}

For the Standard-Model-centered Asimov projection, we define the shift of each pseudo-observable $O_i$ as
\begin{align}
 \Delta O_i(\boldsymbol C) = O_i(\boldsymbol C)-O_i^{\rm SM}.
 \label{eq:app-observable-shift}
\end{align}
Assuming uncorrelated projected uncertainties, the nominal FCC-ee
likelihood is constructed as
\begin{align}
 \Delta\chi^2_{\rm proj}(\boldsymbol C)
 = \sum_i \left[ \frac{\Delta O_i(\boldsymbol C)} {\Delta\sigma_i^{\rm proj}} \right]^2,
 \label{eq:app-future-quadratic-likelihood}
\end{align}
where $\Delta\sigma_i^{\rm proj}$ is the projected experimental uncertainty of pseudo-observable $O_i$, adopted from Ref.~\cite{Allwicher:2023shc}.
The observable shifts are evaluated to linear order in the Wilson coefficients,
\begin{align}
 \Delta O_i(\boldsymbol C) \simeq 
 \sum_a \left. \frac{\partial O_i}{\partial C_a} \right|_{\boldsymbol C=0} C_a.
\end{align}
SM intrinsic and parametric uncertainties are not included in the nominal projection.
The resulting likelihood should therefore be interpreted as a target sensitivity conditional on future theoretical predictions and a consistent pseudo-observable extraction.

In the $(C_{VV}^{b\tau},C_{VV}^{c\tau})$ plane, the projected FCC-ee likelihood is
\begin{align}
 \Delta\chi^2_{\rm FCC\text{-}ee\,EWPO}=\,
 &2.75\times10^5\,(C_{VV}^{b\tau})^2 \label{eq:app_bctau_EWPO}\\
 +&3.18\times10^4\,C_{VV}^{b\tau}C_{VV}^{c\tau}+ 920\,(C_{VV}^{c\tau})^2.\nonumber
\end{align}
The corresponding correlation coefficient is $\rho=-0.9999$, showing that the projected likelihood is nearly, but not exactly, rank one.
The dominant tau-asymmetry response is
\begin{align}
 \Delta A_\tau =\, -0.0232\,C_{VV}^{b\tau} -0.00134\,C_{VV}^{c\tau}.
\end{align}
This observable alone is insensitive along the exact cancellation direction $C_{VV}^{c\tau}\simeq-17.3C_{VV}^{b\tau}$.
Additional heavy-flavor observables make the complete projected matrix formally rank two.

In the $(C_{VV}^{c\tau},C_{VV}^{c\mu})$ plane, the nominal projected likelihood including the effective $A_\tau$ and $A_\mu$ inputs is
\begin{align}
 \Delta\chi^2_{\rm FCC\text{-}ee\,EWPO}=\,
 & 922\,(C_{VV}^{c\tau})^2 -88.0\,C_{VV}^{c\tau}C_{VV}^{c\mu} \nonumber\\
 &+ 3.75\times10^3\,(C_{VV}^{c\mu})^2.
 \label{eq:app_ctaumu_EWPO}
\end{align}
Its correlation coefficient is $\rho=0.024$.
The nominal tau- and muon-asymmetry inputs therefore provide approximately independent response directions in this coefficient plane.

The interpretation of the projected $A_\mu$ input requires care because
$ A_{\rm FB}^{0,\mu} = 3A_eA_\mu/4.$
 The nominal diagonal benchmark does not include the complete covariance among $A_e$, $A_\mu$, and $A_{\rm FB}^{0,\mu}$.
As a robustness test, we remove the separately included $A_{\rm FB}^{0,\mu}$ term while retaining the projected $A_\mu$ information.
The resulting likelihood is very similar to Eq.~\eqref{eq:app_ctaumu_EWPO} and the marginalized $C_{VV}^{c\mu}$ width changes by only $0.12\%$.
It therefore has a negligible numerical impact.

\subsection{Construction of the two-dimensional contours}
\label{app:2D_contours}
The one-parameter four-body reaches are converted into Gaussian coefficient likelihoods before being combined with the projected electroweak information.
For a one-parameter $95\%$ CL scale reach $\Lambda_i$, the corresponding coefficient interval is $ C_{i,95\%} = \Lambda_i^{-2}$.
Using the one-parameter threshold $\Delta\chi^2=3.84$, the diagonal information coefficient is
\begin{align}
 F_{ii} = 3.84\,\Lambda_i^4.
 \label{eq:app_1D_conversion}
\end{align}
The direct flavor categories are treated as statistically independent for simplicity.
Explicitly, for the $b$--$\tau$ versus $c$--$\tau$ plane, the sum of Eq.~\eqref{eq:app_bctau_EWPO} and corresponding sensitivity from 4-fermion processes gives
\begin{align}
 \Delta\chi^2_{\rm FCC\text{-}ee\,combi}=\,
 &2.75\times10^5\,(C_{VV}^{b\tau})^2+3.18\times10^4\,C_{VV}^{b\tau}C_{VV}^{c\tau}\nonumber\\
 +&1.58\times10^3\,(C_{VV}^{c\tau})^2.
\end{align}
The 4-fermion categories add the information along the weak quark-flavor direction of the projected EWPO likelihood.

For the $c$--$\tau$ versus $c$--$\mu$ plane, it gives
\begin{align}
 \Delta\chi^2_{\rm FCC\text{-}ee\,combi}=\,
& 1.59\times10^3\,(C_{VV}^{c\tau})^2-88.0\,C_{VV}^{c\tau}C_{VV}^{c\mu}\nonumber\\
 +&4.30\times10^3\,(C_{VV}^{c\mu})^2.
\end{align}
The nominal EWPO likelihood in this plane is already well conditioned because the projected tau- and muon-asymmetry inputs probe approximately independent lepton-flavor directions.
The four-body information therefore improves both constraints without producing the dramatic rotation found in the $b$--$\tau$ versus $c$--$\tau$ plane.

All two-dimensional $95\%$ CL boundaries displayed in Fig.~\ref{fig:2D} satisfy $\Delta\chi^2 \simeq 5.99$.
This two-parameter threshold differs from the one-parameter value used in Eq.~\eqref{eq:app_1D_conversion} to reconstruct the input likelihoods.
The corresponding scales $\Lambda_i$ characterize the projected coefficient widths of the simultaneous two-parameter $95\%$ confidence region.

For the combined $b$--$\tau$ versus $c$--$\tau$ likelihood, the projected coefficient widths are
\begin{align}
 |C_{VV}^{b\tau}|< 0.0072~{\rm TeV}^{-2},
 \,\,\,\, |C_{VV}^{c\tau}| < 0.095~{\rm TeV}^{-2}.
\end{align}
These widths correspond to
\begin{align}
 \Lambda_{VV}^{b\tau} > 11.8~{\rm TeV},
 \,\,\,\, \Lambda_{VV}^{c\tau} > 3.24~{\rm TeV}.
\end{align}
The large $b$--$\tau$ scale is driven by the precise EWPO direction and becomes individually meaningful only after the flavor-resolved measurements restrict the nearly flat cancellation direction.

For the combined $c$--$\tau$ versus $c$--$\mu$ likelihood, the projected coefficient widths are
\begin{align}
 |C_{VV}^{c\tau}| < 0.0616~{\rm TeV}^{-2},
 \,\,\,\, |C_{VV}^{c\mu}| < 0.0373~{\rm TeV}^{-2}.
\end{align}
These conditions respectively correspond to
\begin{align}
 \Lambda_{VV}^{c\tau} > 4.03~{\rm TeV},
 \,\,\,\, \Lambda_{VV}^{c\mu} > 5.17~{\rm TeV}.
\end{align}
In this plane, the exclusive measurements provide an independent tree-level constraint in addition to the already well-conditioned nominal EWPO likelihood.

The present and projected EWPO likelihoods are alternative scenarios and are never added to one another.
The representative LHC contours in Fig.~\ref{fig:2D} are shown only for present-day orientation and are not included in the quoted FCC-ee combinations.
The construction of the LHC contours is described in Appendix~\ref{app:LHC_comparison}.

\section{Collider sensitivities}
\label{app:LHC_comparison}
The present-collider contours in Fig.~\ref{fig:2D} are included only as reference comparisons with the projected FCC-ee sensitivities.
They are not obtained from official multidimensional likelihoods.
We instead translate representative one-parameter constraints into the coefficient normalization used in this work and reconstruct diagonal Gaussian likelihoods.
The resulting contours should therefore be interpreted as approximate present-sensitivity benchmarks.

For the $b$--$\mu$ interaction, we use the CMS search for non-resonant high-mass dilepton production in association with bottom-tagged jets~\cite{CMS:2025tlo}.
The analysis constrains contact interactions between bottom quarks and charged leptons using the high-mass dimuon spectra in categories with different bottom-tag multiplicities.
The reported contact scales correspond to approximately
\begin{align}
 \Lambda_{b\mu}^{95\%}=2.3~{\rm TeV}.
\end{align}

For the $b$--$\tau$ interaction, we use the high-mass ditau constraint on a third-generation vector-leptoquark benchmark~\cite{ATLAS:2025oiy}.
In the heavy-mediator limit, the relevant left-handed contact scale is approximately given as $ \Lambda_{b\tau}^{95\%} = 1.0~{\rm TeV}$.
For the present orientation benchmark, we assume that the corresponding scale sensitivity is approximately independent of chirality and apply it to the correlated $VV$ direction considered here.

No dedicated official likelihood for the $c$--$\tau$ or $c$--$\mu$ VV direction is used here.
The charm-current entries are estimated from the corresponding high-mass ditau and dimuon sensitivities together with the change in the charm- and bottom-initiated parton luminosities.
This procedure gives the representative values $\Lambda_{c\tau}^{95\%}\simeq1.2~{\rm TeV}$ and $\Lambda_{c\mu}^{95\%}\simeq2.75~{\rm TeV}$.

For visualization, each one-parameter scale is converted into a centered Gaussian coefficient likelihood using the one-parameter criterion $\Delta\chi^2=3.84$.
The resulting approximation is
\begin{align}
 \Delta\chi^2_{\rm LHC} = \sum_i  3.84 \left( \frac{C_i}{\Lambda_i^{-2}} \right)^2.
\end{align}
In the $(C_{VV}^{b\tau},C_{VV}^{c\tau})$ plane, this prescription gives
\begin{align}
 \Delta\chi^2_{\rm LHC}(b\tau,c\tau) = 3.84\,(C_{VV}^{b\tau})^2 +7.97\,(C_{VV}^{c\tau})^2.
\end{align}
In the $(C_{VV}^{c\tau},C_{VV}^{c\mu})$ plane, it gives
\begin{align}
 \Delta\chi^2_{\rm LHC}(c\tau,c\mu) = 7.97\,(C_{VV}^{c\tau})^2 +220\,(C_{VV}^{c\mu})^2.
\end{align}
The displayed LHC boundaries use the common two-parameter criterion $\Delta\chi^2=5.99$ adopted for Fig.~\ref{fig:2D}.
This Gaussian construction neglects sign dependence and correlations between operator directions.
The LHC contours are therefore shown only to indicate the approximate scale of present collider sensitivity.

Flavor observables can provide additional constraints on these interactions, but their interpretation depends on the electroweak embedding and quark-flavor alignment \cite{Allwicher:2023shc}.
Direct lepton-universality tests in bottomonium and charmonium decays currently probe the flavor-diagonal directions only below the TeV scale \cite{Garcia-Duque:2021qmg,Iguro:2024hyk,PDG2026}, while stronger flavor-changing constraints arise only for specific flavor structures.
Such constraints depend on the quark-flavor alignment and do not apply universally to the correlated benchmark directions considered here.
In particular, the choice $C_{\ell q}^{(3)}=0$ removes the direct charged-current component associated with the triplet operator. 
We therefore restrict the quantitative present-day comparison to the more directly applicable LHC bounds.

\bibliographystyle{utphys28mod}
\bibliography{refs}
\end{document}